# Automated Detection of Left Ventricle in Arterial Input Function Images for Inline Perfusion Mapping using Deep Learning: A study of 15,000 Patients

Hui Xue[1], Ethan Tseng[1], Kristopher D Knott[2], Tushar Kotecha[3], Louise Brown[4], Sven Plein[4], Marianna Fontana[3], James C Moon[2], Peter Kellman[1]

1. National Heart, Lung and Blood Institute, National Institutes of Health, Bethesda, MD, USA
2. Barts Heart Centre, London, UK
3. National Amyloidosis Centre, Royal Free Hospital, London, UK
4. Department of Biomedical Imaging Science, Leeds Institute of Cardiovascular and Metabolic Medicine, University of Leeds, Leeds, UK

## Corresponding author:

Hui Xue

National Heart, Lung and Blood Institute
National Institutes of Health
10 Center Drive, Bethesda
MD 20892
USA

Phone:  +1 (301) 827-0156
Cell:   +1 (609) 712-3398
Fax:    +1 (301) 496-2389
Email: hui.xue@nih.gov

**Word Count:** 4,838

hui.xue@nih.gov
ethan.tseng@yahoo.com
kristopher.knott@nhs.net
tushar.kotecha@nhs.net
L.Brown1@leeds.ac.uk
S.Plein@leeds.ac.uk
m.fontana@ucl.ac.uk
james@moonmail.co.uk
kellmanp@nhlbi.nih.gov

**Abstract**

**Purpose**

Quantification of myocardial perfusion has the potential to improve the detection of regional and global flow reduction. Significant effort has been made to automate the workflow, where one essential step is the arterial input function (AIF) extraction. Failure to accurately identify the left ventricle (LV) prevents AIF estimation required for quantification, thus high detection accuracy is required. This study presents a robust LV detection method using the convolutional neural network (CNN).

**Methods**

CNN models were trained by assembling 25,027 scans (N=12,984 patients) from three hospitals, seven scanners. Performance was evaluated using a hold-out test set of 5,721 scans (N=2,805 patients). Model inputs were a time series of AIF images (2D+T). Two variations were investigated: a) Two Classes (2CS) for background and foreground (LV mask); b) Three Classes (3CS) for background, LV and RV. The final model was deployed on MRI scanners using the Gadgetron reconstruction software framework.

**Results**

Model loading on the MRI scanner took ~340ms and applying the model took ~180ms. The 3CS model successfully detected the LV in 99.98% of all test cases (1 failure out of 5,721). The mean Dice ratio for 3CS was 0.87±0.08 with 92.0% of all cases having Dice >0.75. The 2CS model gave a lower Dice ratio of 0.82±0.22 (P<1e-5). There was no significant difference in foot-time, peak-time, first-pass duration, peak value and area-under-curve (P>0.2) comparing automatically extracted AIF signals with signals from manually drawn contours.

**Conclusions**

A CNN based solution to detect the LV blood pool from the arterial input function image series was developed, validated and deployed. A high LV detection accuracy of 99.98% was achieved.

# Introduction

Quantification of myocardial perfusion using dynamic cardiovascular magnetic resonance has been studied for over 20 years (1–7). Perfusion quantification has potential to improve disease detection involving global flow reduction such as balanced multi-vessel coronary stenosis or microvascular dysfunction (8,9) and has been shown to have prognostic significance (10).

Significant effort has been made to automate the workflow of perfusion quantification, including imaging sequences (11,12), AIF signal correction (13–16), motion correction (17–20), and sector or pixel-wise perfusion estimation using either deconvolution (2,4,21) or tissue kinetics model (5,22–25). Deconvolution methods include exponential decay (4), Fermi function (2,26) and BSpline response function (27). Tissue kinetics methods formulate the myocardial Gd transport with partial differential equations (28,29) and explicitly quantify the Gd extraction into the interstitium.

One automated perfusion quantification system (5,30) was developed by our group. This solution utilizes a "dual-sequence" perfusion imaging scheme with a dual-echo AIF for T2* estimation. Perfusion images are acquired while the patient breathes freely and automated motion correction is applied to align the myocardium over multiple heart beats, which enables pixel-wise perfusion mapping. Since the accuracy of perfusion quantification highly depends on the correct measurement of the input contrast concentration, the "dual-sequence" imaging method acquires AIF images with low spatial resolution, very short saturation time and echo time, to reduce signal saturation caused by high contrast concentration in the blood.

Automated extraction of the AIF signal during contrast bolus passage is a prerequisite for automated perfusion quantification (Figure 1). An "ad-hoc" algorithm was previously proposed in (30) to detect a binary mask of the left ventricle from AIF image series (described in section 2.4 of (30)). This algorithm was based on simple thresholding to mask off background and non-enhanced tissues within the field-of-view based on the signal up-slope. Resulting pixels were further classified into LV and RV based on the k-means algorithm with 2 initial clusters. The LV cluster was refined using the k-means algorithm repeated with 4 clusters to select LV blood-pool pixels. The AIF signal was computed by keeping pixels in the LV cluster that had high peak intensity in order to minimize partial-volume effects. While this method worked well during our initial testing (30), a higher than desired failure rate (defined by failing to detect the LV) was revealed after application to larger numbers of clinical adenosine stress studies. This occurred especially for patients with reduced or delayed AIF contrast concentration, possibly due to low cardiac output or incomplete contrast injection. Another reported method (31) to automatically detect AIF LV was based on thresholding the standard deviation map of time series, followed by LV pixel masking using independent component analysis. The AIF mask was also eroded to select high intensity pixels for time-intensity curve. While both algorithms utilized the contrast uptake dynamics of the LV, they were ad-hoc using manually crafted features (e.g. standard deviation or upslope maps) designed for key algorithmic components, together with inflexible thresholding or clustering. Alternatively, a deep learning based solution has the potential to overcome these limitations by learning what is important for detecting the LV without potentially brittle "feature engineering" (32).

Failed detection of the AIF LV will result in incorrect perfusion mapping. High LV detection accuracy is required for an automated solution. For this purpose, this study proposed an AIF LV detection method using deep learning. A convolutional neural net model was trained by assembling a very large data cohort from three hospitals. The 2D+T temporal AIF image series was input into CNNs. A total of 25,027 scans comprised of both stress and rest scans from N=12,984 patients were collected for training. To test the trained model, an independent hold-out test set was collected, including 5,721 perfusion scans from N=2,805 patients. After training and evaluation, the final CNN model was integrated into the automated perfusion mapping software that was deployed on MRI scanners via the Gadgetron InlineAI (33) framework. The pre-trained CNN model was applied to the incoming AIF image series and the LV mask was generated as output.

## Methods

*Imaging and data collection*

Three hospitals participated in this study (Barts Heart Centre, Barts; Royal Free Hospital, RFH; Leeds Teaching Hospitals NHS Trust, LTHT). A total of seven MRI scanners were used. Five scanners were at Barts Heart Centre, including three 1.5T MAGNETOM Aera and two MAGNETOM Prisma, Siemens AG Healthcare, Erlangen, Germany. A 1.5T MAGNETOM Aera was used at the Royal Free Hospital and LTHT had a MAGNETOM Prisma 3T scanner.

Adenosine stress and rest perfusion imaging was performed at participating sites using the "dual-sequence" scheme (5). This imaging sequence utilized saturation recovery and single-shot readout. Three short axis slices were imaged typically for 60 heart beats after injection of the contrast bolus. An AIF imaging module was inserted before the perfusion imaging and played out after the R wave trigger with very short delay. AIF images were acquired at the most basal slice. Imaging parameters for AIF were: FOV 360×270 mm$^2$, slice thickness 10 mm, small imaging matrix 64×48 (linear order), interleaved acceleration R=3, two echo readouts, TE=0.76 and 1.76ms, TR=2.45ms, short trigger delay TD=4.2ms, imaging duration 42ms, flip angle 5º, FLASH readout, total duration including saturation recovery preparation was 57ms. The same imaging sequence was used in all hospitals for all patients.

Data were acquired with the required ethical and/or audit secondary use approvals or guidelines (as per each center) that permitted retrospective analysis of anonymized data for the purpose of technical development, protocol optimization and quality control. All data were anonymized and de-linked for analysis by NIH with approval by the NIH Office of Human Subjects Research OHSR (Exemption #13156). A total of 25,027 scans (10,349 stress) were retrospectively included for training the model, which counted for N=12,984 consecutive patients at three sites (941 patients had only rest perfusion). This data cohort was acquired from 29 April 2016 to 15 February 2019. To test the AIF detection, a separate test set was assembled retrospectively. Imaging for the test set was from 16 February 2019 to 13 August 2019. This cohort consists of N=2,805 patients and 5,721 perfusion scans (2,780 stress).

*Neural Net model*

The AIF image series was first temporally resampled to be on a 0.5s grid (5,30), to account for heart rate variation or any missed ECG triggers during image acquisition. The first 64 saturation recovery images (initial proton density images are excluded) after temporal resampling were selected, resulting in an image array of 64×48×64. This image array was used for training and testing. Figure 1 gives an example of an AIF image series (first 48 of 64 saturation recovery images) after resampling. The corresponding LV mask is given in Figure 2(a). Figure 2(b) illustrate the temporal signal with foot, peak, and valley times annotated.

The U-net semantic segmentation architecture (34,35) was modified for this application of AIF LV detection. As shown in Figure 3, a ResNet module (36) was used as the basic building block, which consisted of: batch normalization (BN), rectified linear unit (Relu), and convolution (CONV) components as shown. To simplify the neural net specification, one module consists of two convolution layers with the same number of filter outputs. As shown in Figure 3, the downsampling and upsampling branches were connected by "skip-connections". The AIF times series went through the input layer followed by the downsampling branch. As the spatial resolution was reduced, the number of filters increased. The upsampling branch increased the spatial resolution and reduced the number of filters. Final CONV layer output a 64×48×C array for segmented classes C (either 2 or 3). The output of the NN was input into a pixel-wise sigmoid or softmax layer to estimate probability maps. Final segmentation masks were derived by thresholding the probability maps. All CONV layers used a 3×3 kernel with stride 1 and padding 1. To utilize the temporal information in the dynamic AIF image series, the input layer in the implemented neural network received the 2D+T time series with the temporal dimension as input channels. The input array of size 64×48×64 was convolved spatially with a total of 96 convolution kernels. Each kernel has the size of 3×3. Convolution was not used for the temporal dimension; instead, a full dense connection was applied to 64 input channels. To better illustrate this point, consider a single pixel in the AIF image: it will be a vector $x$ of length 64. The full connection along the temporal dimension is equivalent to having a dense parameter matrix $M \in R^{96 \times 64}$. This matrix is applied to the input vector (length 64) and produces an output response $z$ of length 96:

$$\begin{bmatrix} | \\ z \\ | \end{bmatrix} = \begin{bmatrix} m_{1,1} & m_{1,2} & m_{1,3} & \ldots & m_{1,63} & m_{1,64} \\ m_{2,1} & m_{2,2} & m_{2,3} & \ldots & m_{2,63} & m_{2,64} \\ m_{3,1} & m_{3,2} & m_{3,3} & \ldots & m_{3,63} & m_{3,64} \\ \ldots & \ldots & \ldots & \ldots & \ldots & \ldots \\ m_{95,1} & m_{95,2} & m_{95,3} & \ldots & m_{95,63} & m_{95,64} \\ m_{96,1} & m_{96,2} & m_{96,3} & \ldots & m_{96,63} & m_{96,64} \end{bmatrix} \begin{bmatrix} | \\ x \\ | \end{bmatrix} \qquad (1)$$

The first convolutional network layer (in Figure 3) outputs a 64×48×96 array which then went through batch normalization and a nonlinear layer. The layer parameters were optimized during backpropagation to refine their abilities to capture contrast dynamics. As a result, the segmentation accuracy was gradually increased during training.

The loss function was a weighted sum of cross-entropy and the Intersection Over Union (so-called IoU or Jaccard index). For the binary mask $y^i$ and predicted probability $\hat{y}^i$, the IoU was computed as:

$$J_{IoU} = \frac{1}{N} \sum_{i=0}^{N-1} \frac{y^i \hat{y}^i}{y^i + \hat{y}^i - y^i \hat{y}^i} \qquad (2)$$

where N is the number of pixels and $i$ is the index of pixels. The total cost was the weighted sum (37):

$$J = J_{cross-entropy} - 0.5 \cdot \log(J_{IoU}) \qquad (3)$$

This cost function simultaneously maximizes the probability that a pixel is correctly classified and optimizes the overlap between the detected mask and ground-truth. This scheme has demonstrated improved segmentation accuracy in past MICCAI segmentation challenge (38).

Each AIF dataset was corrected for spatial variation in receive coil sensitivities using the proton density weighted images acquired at the start of acquisition, and were scaled to have the largest value be 1 before input to training or testing: $AIF = AIF/max(AIF)$. Random flipping along readout and phase directions (probability of apply flipping was 0.5) was used as a data augmentation step before training.

*Bootstrap data labelling*

Given the large number of datasets, a bootstrap strategy was implemented to speed up data labelling. First, the respiratory motion was corrected from AIF image series using a motion correction algorithm designed for free-breathing cardiac perfusion (5,30). The motion correction algorithm was designed for robustness in the presence of contrast variation by decoupling the respiratory motion and image contrast variation due to bolus passage using the temporal Karhunen-Loève (KL) transform. The motion correction starts with large temporal window and gradually refines the estimated deformation field by reducing the temporal window. The motion corrected images series were input for LV labelling using the "ad-hoc" algorithm previously proposed (5,30). In this method, the noise background was removed using a simple thresholding with SNR < 3. The LV was detected by clustering the temporal intensity curves. This heuristic algorithm was applied to all datasets. All results were visually inspected to either accept or correct tentative labels (H.X., with 9 years of experiences in perfusion imaging). Failed cases were manually corrected and included in the training set. Approx. 200 hours were spent on labelling the entire data cohort.

*LV detection utilizing 2D+T contrast dynamics*

A 2D+T temporal image array of motion corrected AIF images was assembled and input into the CNN. To train the CNN with more input features, the right ventricle was further marked as another object class. The motivation was to utilize the unique contrast update dynamics and consistent anatomy between the LV and RV, since the RV enhances prior to the arrival of contrast in the LV. Two variations of NNs were investigated: a) Two Classes (2CS) for background and foreground (LV mask); b) Three Classes (3CS) for background, foreground LV and foreground RV regions. It was hypothesized that the extra information on RV contrast dynamics and spatial relationship to LV may lead to improved LV detection.

*Training and hyperparameter search*

The training data were randomly shuffled and split into training set (Tra) (number of scans, N=22,941) and development set (Dev, N=2,086). Figure 4(a) illustrates the large variation in patient anatomy and image prescriptions of the AIF image series. Furthermore, there is considerable variation of the AIF temporal response, in both bolus arrival time and duration (Figure 4(b)). The CNN model was implemented using PyTorch (39) and training was performed on an Ubuntu 18.04 PC with four NVIDIA GTX 2080Ti GPU cards, each with 11GB RAM. In initial experiments, it was determined that the number of ResNet modules and the number of CONV filters in the NN had the most significant impact on detection accuracy. Therefore, a hyperparameter search was conducted to test different combinations (5 to 9 ResNet modules and number of CONV filters from 64 to 256). For all tests, the Adam optimizer was used with initial learning rate of 0.001. The betas were 0.9 and 0.999 and epsilon was 1e-8. Learning rate was reduced by x2 for every 10 epochs. Training took 40 epochs and the best model was selected as the one giving the best performance on the Dev set. The training process was repeated with a randomly chosen smaller training set of 1,500 samples, to investigate the performance for a smaller training dataset size. The resulting models were tested on the same hold-out test set.

*Analysis*

The success of LV detection was initially quantified by Dice ratio (40). Following convention often used in deep learning based object detection and semantic segmentation (41,42), if the Dice ratio between ground-truth mask and NN result was greater than 0.5, the detection was defined as being successful. A Dice ratio over 0.75 was also reported as the second indicator.

The detected AIF signal curve was further compared to the corresponding ground-truth. AIF foot time, peak time, peak value, first-pass duration (from foot to valley, as illustrated in Figure 2(b)), area-under-curve (AUC) for first pass and correlation coefficient (CC) were computed for both auto and manual curves. Since the dual sequence has a low spatial resolution for AIF images, edge pixels of the AIF LV can often have reduced intensity due to partial volume with adjacent tissue. The LV mask was further eroded to keep the top 15% percentile values for computing the time-intensity curves, as in previous studies (30), calculated at peak LV enhancement. The resultant final mask was applied for all AIF images. Selected pixels were averaged to compute the mean AIF signal for each time frame.

The resulting values were presented as mean +/- standard deviation. A t-test was performed and a P-value less than 0.05 was considered statistically significant.

*Inline integration*

Training was conducted offline using Pytorch. The resulting models were integrated to run inline on MRI scanners. The integration was achieved using the Gadgetron Inline AI (33). The key components were a set of interfaces to load pre-saved NN models and apply the model to incoming data (so-called "inference"). This involved the transfer of model objects from Pytorch to C++ and passing data from C++ to Pytorch modules. Since Gadgetron may run with different computing configurations, such as on the scanner image reconstruction computer, a user supplied computing server, or on cloud nodes (43,44), the model inference step was chosen to run without GPU. That is, the inference was performed on CPU. While training can be prohibitively slow with only CPU, our tests showed CPU inference for AIF detection was sub-second for clinical perfusion scans.

The NN based AIF detection was integrated into inline perfusion mapping workflow. The model loading was triggered as soon as the imaging sequence was initialized and performed in parallel with data acquisition. It did not increase the overall reconstruction time. The loaded model was kept in the Gadgetron runtime environment and applied to the incoming AIF image array. The resulting LV mask and AIF signal were used for downstream perfusion quantification. Thus, using the Gadgetron streaming framework with the Inline AI infrastructure, an AIF detection scheme powered by deep learning was achieved on clinical MRI scanners.

## Results

A training session took ~40 mins for 40 epochs. Figure 5(a) gives the cost vs. the number of iterations in training. The Adam optimizer used in the training was very efficient in driving down the total cost and increasing the detection accuracy. A total of 45 training sessions was performed for the hyperparameter search, which took ~32 hours. The best performance achieved using a hyperparameter search used an architecture consisting of 8 ResNet modules. Each module used 128 CONV filters. These parameters were selected for all following performance measurements.

The three-class segmentation outperformed two-class segmentation. The 3CS model successfully detected the LV for 99.98% of all test set cases (1 failed out of 5,721 cases), while the original ad-hoc detection had 52 failed cases. The mean Dice ratios for 3CS and 2CS were 0.87±0.08 and 0.82±0.22 ($P<1e-5$), respectively. Figure 5(b) gives the histogram of three-class detection for all test cases showing that 92.0% of all test cases had a Dice ratio higher than 0.75. As a comparison, the two-class detection method had only 80.8% cases with a Dice ratio higher than 0.75.

Figure 6 gives examples of AIF detection with different Dice ratios from the Test set, to illustrate detection performance for different levels of accuracy. Those cases with lower Dice still located the LV blood pool successfully.

The correlation coefficient (CC) between automated and ground-truth AIF signals was 0.998±0.012. Figure 7 gives the Bland-Altman plots of NN results vs. manually labelled masking. Foot and peak time for manually labelled data were 14.0±3.41s and 18.0±4.52s, compared to three-class NN of 14.0±3.42s and 18.0±4.55s. First-pass duration was 13.2±4.97s and 13.2±5.00s for manual label and NN, respectively. Normalized intensity values at peak were 0.385±0.099 for manual label and 0.383±0.099 for NN. AUC was 6.14±2.18 for manual label and 6.12±2.19 for NN. No significant differences were found for all results (P>0.2). Mean differences in percentage for foot and peak time, peak intensity values and AUC was 0.51%, 0.52%, 1.61% and 2.00% respectively.

Training was repeated with 1,500 training samples and the resulting model was applied to the same test cohort (N=5,721). With the more limited size training dataset, the Dice ratio was reduced to 0.81±0.14, with 188 test cases that had a Dice ratio lower than 0.5 (3.3% failure rate). These 188 cases were visually inspected and it was determined that 29 cases incorrectly selected the RV and 33 cases selected regions outside heart.

The NN model was integrated and tested on all MRI scanners as a component of inline perfusion mapping. The reconstruction workflow was configured so that the Gadgetron server started to load the saved model while a perfusion sequence was under preparation on the scanner. Model loading time was ~340ms and applying the model to an incoming AIF image array was ~180ms. These timing results were measured with CPU inference (2× Intel Xeon Gold 6152 CPU, 2.10GHz, 192GB RAM).

## Discussion

Reliability is paramount for in-line mapping solutions since they must be fully automatic. The proposed CNN based solution greatly improved on the prior method by approximately 50:1, decreasing the failure rate from 0.9% to less than 0.02%. High reliability is particularly important for contrast enhanced stress tests, since it is difficult to repeat the acquisition. The motivation to achieve a very low failure rate in this study was to minimize introducing failures in the flow mapping process, which will be added on top of other possible error sources in acquisition. Although other factors such as poor ECG triggering can lead to problems, these can be minimal at experienced sites that do stress studies on a routine basis. Poor ECG triggering and other problems may be resolved prior to administering stress with exceptions such as some types of arrhythmias. The failure of automated in-line processing to correctly detect the LV would lead to a full failure of the pixel-wise flow mapping which would be unacceptable even at a 1% failure rate, requiring a repeated stress study. The automated process should not increase the apparent failure rate for stress studies.

It was shown that high accuracy was achieved when training on a large dataset that captured sufficient variation, and that training on a limited size dataset greatly reduced the performance. Variation included both spatial and temporal characteristics for a range of patient positions, orientations, and sizes, as well as time delay of enhancement and bolus duration. The performance of the CNN with small training size was actually worse than the original ad-hoc method. Employing a small training dataset size risks over-fitting. It has been recognized that the performance of deep neural networks increases with larger datasets (45). From our experience to conduct deep learning development, a major effort is required for manual labelling thus we used a bootstrap approach as a practical means of efficiently labelling a large training dataset. Recent studies explored effective learning strategies if large datasets are not available (e.g. fewer-shot learning (46)) and effective methods to improve data labelling process (e.g. weak supervision (47)). Their applications to clinical imaging AI are of high interest.

The proposed AI method for LV detection had a single failure with DICE = 0, where the LV was incorrectly identified as the RV. This case is illustrated in Supporting Information

Figure S1. In this case, the bolus arrived at the RV very late and, the LV blood pool did not enhance at all during the duration of the imaging. Thus, the perfusion mapping would not have succeeded even with correctly identified LV.

While no prior study has reported on the application of a convolutional neural net for perfusion AIF LV detection, pixel-wise image segmentation using NN has been well studied. The first well-known algorithm is the fully convolutional network (FCN) (48), which repurposed classification networks for pixel-wise segmentation by modifying the output layer and loss function. Extensions of these networks include adding encoding-decoding structures and expanding the filter depth while reducing the spatial resolution (34,49–51). These NNs also added skip connections between corresponding down/upsampling layers, to allow detailed pixel-level segmentation. Motivation to use skip connections include allowing image information to pass between different spatial resolution levels and to help overcome the vanishing gradient problem during optimization. These connections allowed convolution layers at a spatial resolution level to receive information from other spatial resolution levels, favoring parameter optimization that captures anatomical structures at different sizes or scales. Otherwise, the downsampling step may lead to loss of some spatial detail in the final segmentation. During the backpropagation, these connections also allowed the gradient to pass as a shortcut and thus speed up the effective updating of network parameters. These have led to major improvement in segmentation performance, as demonstrated by U-net (34) and Seg-net (50). These CNN architectures can accept multiple input channels to segment input images. The U-net architecture was further extended by using recurrent (52) and residual modules (53). More recent variations include V-net (49) and TernausNet (37) which kept the skip connections but have a different number of convolution filters and kernel size in each down/upsampling layer. These segmentation NNs have been used for cardiac MRI image segmentation. The first well-known work utilized a modified FCN network to segment myocardium from cine imaging (54). Recent published work includes automated analysis of T1 mapping image (55) and cardiac flow images (56).

The proposed approach utilized a 2D+T AIF image series as input to the NNs, different from cine deep learning segmentation using individual 2D images (54). Since perfusion images show significant contrast variation during bolus passage, inputting the temporal image series allowed the NNs to utilize the full bolus passage information for LV detection. As a pre-processing step, motion correction was applied to AIF image series to correct the respiratory motion from free-breathing imaging. This step allowed a single LV mask to be detected. It also simplified the neural net architecture for LV detection, because omitting MOCO will force NNs to produce a mask for every AIF image. To visualize the characteristics of trained CNNs, the saliency map (57) was computed as the derivative of the CNN loss function to an input AIF image array, as illustrated in Supporting Information Figure S2. Higher magnitude in saliency map indicates corresponding spatial and/or temporal regions in AIF 2D+T array that had greater effect on the CNN loss. In other words, the CNN was trained to "pay attention" to both ventricles during the contrast bolus passage.

Experimental results showed the neural net with three-class training set (3CS) outperformed two-class detection (2CS). This result demonstrated the added value of extra labelling on the given AIF time series. By providing the context information of the RV, the NNs learned the relative LV anatomy better. The contrast enhancement pattern of the RV, which reaches peak contrast sooner than the LV, is a useful prior for NNs to capture. Similar performance boosting by combining additional prior information has been previously described in other applications, such as multi-organ segmentation using CNN (58).

One major barrier for more detailed labelling in deep learning is establishing large training sets, which comes with the need for both expert hours and high financial costs. The computer vision community generates gigantic datasets of labelled images by hiring "common" workers through the internet (e.g. "crowdsourcing" in Amazon Market Place), such as in ImageNet (59) and Microsoft COCO (60). However, this strategy may not be feasible for medical imaging, due to required professional knowledge and expertise. While some methods such as MOCO or the bootstrap strategy used in this study may be helpful to speed up the labelling process, developers of cardiac deep learning applications still spend significant hours to establish labelled training sets (61). Recent development of synthetic data labelling can learn a generative model from a small training set and synthesize more training data by sampling the learned probability distribution (62). This strategy is promising although also requires expert supervision to verify the generated labels are correct. Another related method to mitigate data labelling workload is through transfer learning. As an example, large cardiac cine datasets labelled by experts do exist (e.g. UK biobank data used in (54)). Since early layers of NNs often learn similar low-level features, it is possible to use cine model weights of early layers for another application. In this way less new data may be required.

This paper proposed a solution to directly detect AIF LV blood pool for perfusion mapping. Another possible approach for AIF LV blood pool detection is to first segment the myocardium from high resolution perfusion images and use the segmented LV to help AIF blood pool detection for the dual-sequence image acquisition. Two recent studies (63,64) have proposed and validated convolution neural network based solution for perfusion myocardium segmentation. Neither study has reported using results of myocardial segmentation to segmenting an AIF image series. This study decoupled the AIF detection from the myocardial segmentation, because large datasets were available for AIF detection (as used in this study) by using bootstrap approach with the initial ad-hoc method. The published perfusion myocardial segmentation studies, on the contrary, used smaller datasets (e.g. 175 patients in (63) and 1,034 patients in (64)). An advantage of this decoupling is the accuracy of AIF detection can be independently optimized and evaluated.

This study presented a CNN based solution for AIF LV detection and the inline integration of models on MRI scanners through the Gadgetron InlineAI (33). Although NN training is known to be computationally heavy, we showed that model inference is fast enough to work on a CPU on the MRI scanner. As a result, applying deep learning for advanced cardiac MRI is becoming practical in the clinical setting such as demonstrated by this study.

This study was conducted on the Siemens scanners at three hospitals, using the open-source Gadgetron software platform. It is possible to extend the AIF LV detection and the inline perfusion mapping to other vendors which would require "dual-sequence" perfusion imaging. The Gadgetron framework software is open-source including data conversion software for major vendors, available on the github repository (https://github.com/ismrmrd). The proposed neural network source code and related training scripts are open-source at github repository (https://github.com/xueh2/QPerf). The PyTorch source code is vendor neutral and can be used to train models for different imaging protocols or sequences.

The proposed workflow has the limitation of requiring good motion correction of AIF image series. Failed motion correction degrades the AIF time signal extracted from the temporal image series which affects the quantitative perfusion mapping. Another alternative is to perform per-frame by frame segmentation on the free-breathing image series, which requires significantly more effort in data labelling. Networks also must capture both respiratory motion and contrast passage. This technical route was not explored in this study and remains an open topic. Another limitation concerns the U-Net architecture which is not designed to extract temporal features by convolution (i.e. a translationally invariant manner). However, the training with a large number of studies has resulted in a network that is robust to variations in the temporal characteristics of the AIF.

## Conclusions

This study developed, validated and deployed a CNN based solution to detect the LV blood pool from arterial input function image series. Training was performed on a large set of 25,027 perfusion scans and the resulting model was tested on an additional 5,721 scans. A high LV detection accuracy of 99.98% was achieved with mean Dice ratio 0.87±0.08 for the test dataset.

**Abbreviations**

| | |
|---|---|
| AIF | arterial input function |
| AUC | area-under-curve |
| DSC | Dice similarity coefficient |
| FLASH | fast low angle shot |
| IoU | Intersection over Union |
| MBF | myocardial blood flow |
| MOCO | motion correction |
| CNN | convolutional neural network |

**Declarations**

**Ethical Approval and Consent to participate**

Data was acquired with the required ethical and/or audit secondary use approvals or guidelines (as per each center) that permitted retrospective analysis of anonymized data for the purpose of technical development and protocol optimization and quality control. All data was anonymized and de-linked for analysis by NIH with approval by the NIH Office of Human Subjects Research OHSR (Exemption #13156).

**Consent for publication**

Data was acquired with the required ethical and/or audit secondary use approvals or guidelines (as per each center) that permitted retrospective analysis of anonymized data and publication. The per center IRB approval document is available for review by the Editor-in-Chief.

**Availability of data and material**

The raw data that support the findings of this study are available from the corresponding author upon reasonable request subject to restriction on use by the Office of Human Subjects Research. The source file to train the CNN model is available at https://github.com/xueh2/QPerf.git. Example datasets were provided at this website with a user guide to run the pre-trained models for LV detection. A large dataset of AIF signals for N=1,500 perfusion scans was further shared and can be downloaded from the above-mentioned github repository. All AIF signals were stored as NumPy (.npy) files directly readable in Matlab and Python. The AIF signals have been temporally resampled to have uniform 0.5 second spacing. The file suffix indicated stress ("S") or rest ("R") acquisitions.

**Competing interests**

The authors declare that they have no competing interests.

**Funding**

Supported by the National Heart, Lung and Blood Institute, National Institutes of Health by the Division of Intramural Research and the British Heart Foundation (CH/16/2/32089).24

**Authors' contributions**

HX and PK conceived of the study and drafted the manuscript. HX, PK and ET developed the algorithms, implemented the inline integration of neural net model and performed processing and analysis. KK, TK, LB, SP, MF, JM were responsible for CMR studies used in training and test. All authors participated in revising the manuscript and read and approved the final manuscript.

**List of Captions:**

**Figure 1** Example of AIF image series: 2D+T time series demonstrating the passage of contrast bolus.

**Figure 2** Example of AIF image series: (a) AIF image with LV mask overlaid, (b) AIF time intensity curve extracted from AIF images using the LV mask with the foot, peak and valley points indicated.

**Figure 3** ResUnet architecture used for AIF LV mask detection. The 2D+T AIF time series after motion correction is input to the CNN and passed through the downsample branch with gradually reduced spatial resolution. This process is passed through the upsample branch until the original spatial resolution is restored. Skip-connections between down- and up-sample branches allows the neural net to learn detailed features at each spatial resolution. More ResNet modules can be inserted into the neural net to adjust its depth. A threshold is applied to the probability map to obtain the final LV mask.

**Figure 4** Illustration of variation in AIF: (a) anatomical variation in the data cohort illustrating a range of variation in anatomy and imaging for which the detection algorithm must cope. These AIF images were acquired at the basal slice location. For each time series, the frame shown here was at the approx. moment of peak LV contrast. (b) temporal variation in AIF delay, duration, and amplitude for 3 patients.

**Figure 5** AIF detection performance on the test set. (a) loss vs. the number of iterations for a typical training session, (b) histogram of three-class detection for all test cases, showing majority of cases have high dice scores.

**Figure 6** Examples of AIF detection with different level of performance. The boundary contours of ground-truth manually labelled masks were extracted and plotted in red for comparison purpose. Cases with lower score still located the LV blood pool successfully.

**Figure 7** Bland Altman plots for parameters derived from the AIF signal curve comparing manually labelled vs. CNN. Both bias (dotted line in blue) and 95% limits of agreement (dotted lines in red) are plotted.

**Additional files**

**Supporting Information Figure S1**: The AIF signal for the single failed case shows that the contrast did not reach the LV. (a) AIF time series is presented for all 64 frames. In this case, the contrast arrived very late and the LV did not enhance. (b) Manual labels for LV and RV are overlaid on a frame. (c) Time-intensity profiles for manually labelled ROIs show very late contrast arrival. (d) Neural network model mistakenly detected the RV.

**Supporting Information Figure S2**: The saliency map was computed for an AIF image array to investigate the characteristics of a trained CNN. The saliency map was the derivative of the loss function for an input image array. Higher magnitude in the saliency map indicates corresponding spatial and/or temporal regions in the AIF 2D+T array that had greater effects on the CNN loss.

**Figure 1** Example of AIF image series: 2D+T time series, demonstrating the passage of contrast bolus.

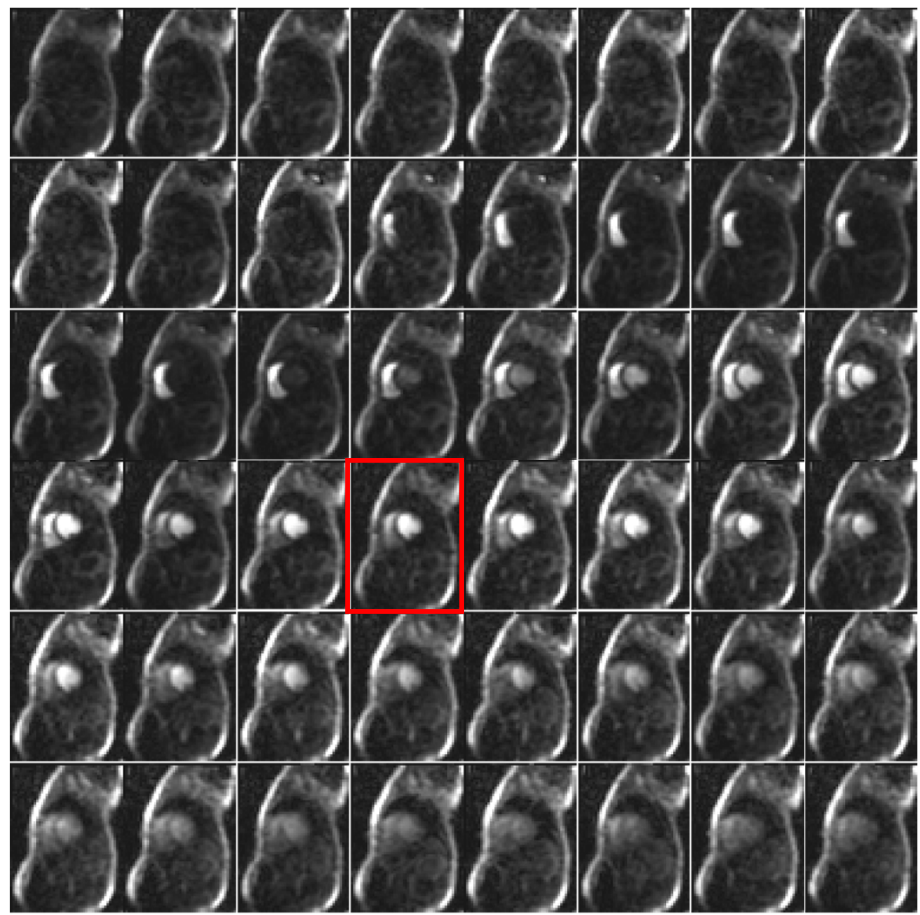

**Figure 2** Example of AIF image series: (a) AIF image with LV mask overlaid. (b) AIF time intensity curve extracted from LV mask with the foot, peak and valley points indicated.

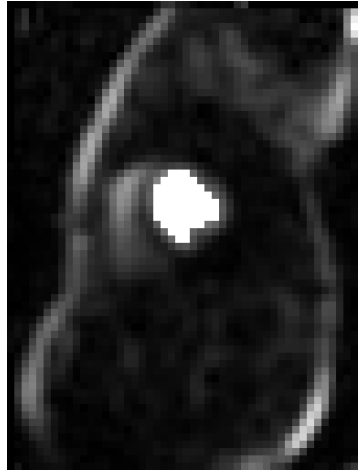
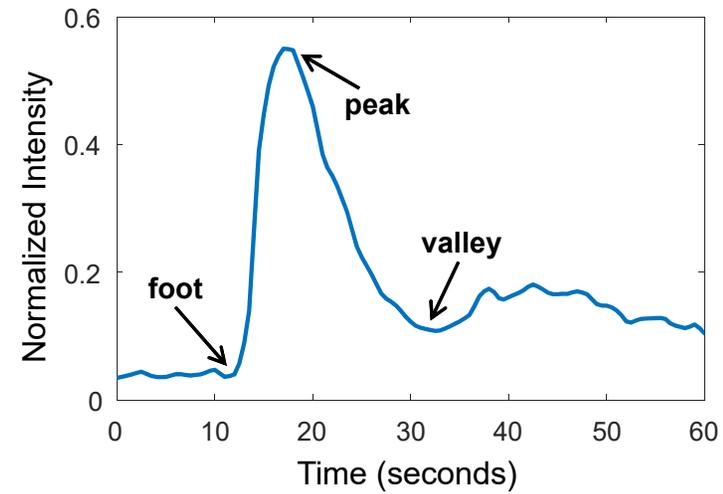

(a) An AIF image with LV mask overlay.   (b) AIF time intensity curve extracted from LV mask.

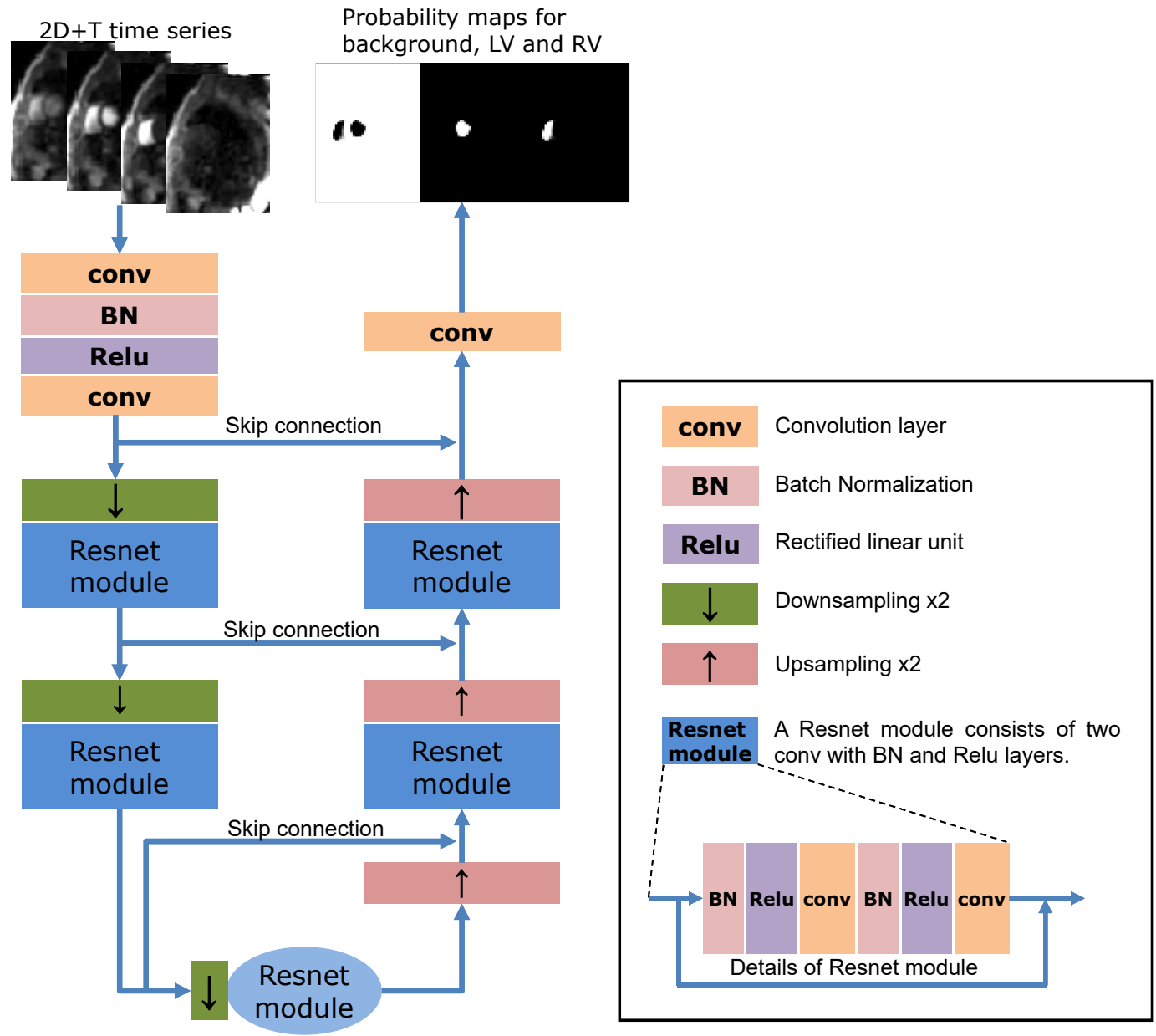

**Figure 3** ResUnet architecture used for AIF LV mask detection. The 2D+T AIF time series after motion correction is input to the CNN and passed through the downsample branch with gradually reduced spatial resolution. This process is reverted through the upsample branch until the original spatial resolution is restored. Skip-connections are established between down- and upsample branches to allow neural net to be adaptively learn through spatial resolution. More resnet modules can be inserted into the neural net to adjust its depth.

**Figure 3** Illustration of anatomical variation in the data cohort illustrating a range of variation in anatomy and imaging for which the detection algorithm must cope. These AIF images were acquired at the basal slice location. For every time series, the frame shown here was at approx. moment with peak LV contrast.

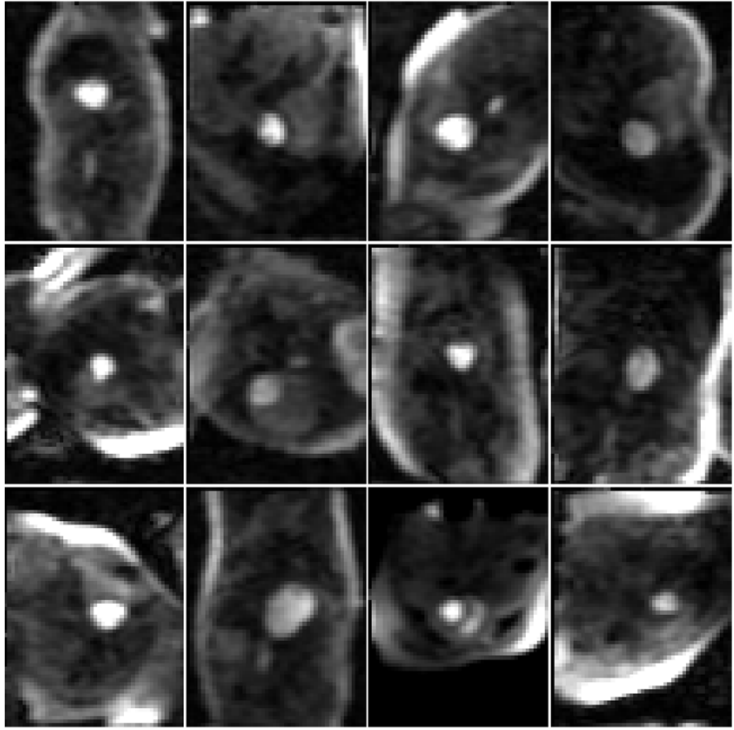

(a) Examples of AIF images of LV enhancement

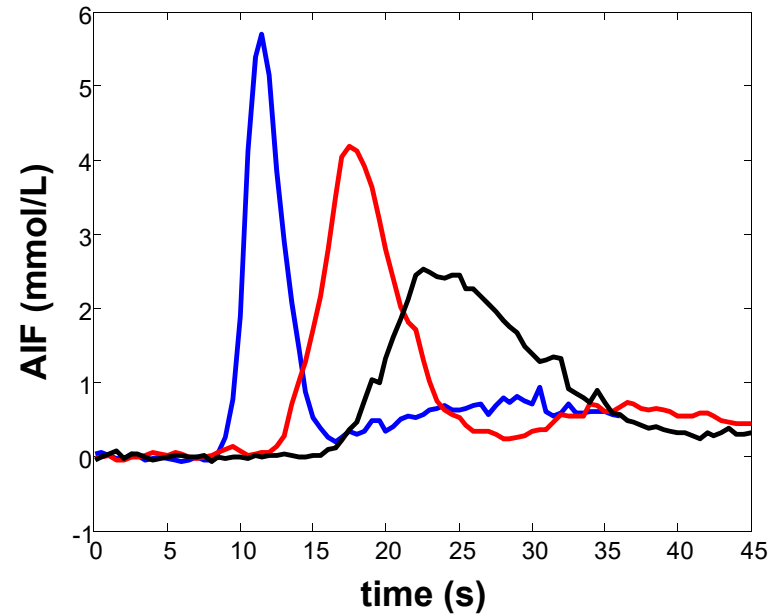

(b) Examples of AIF time signals

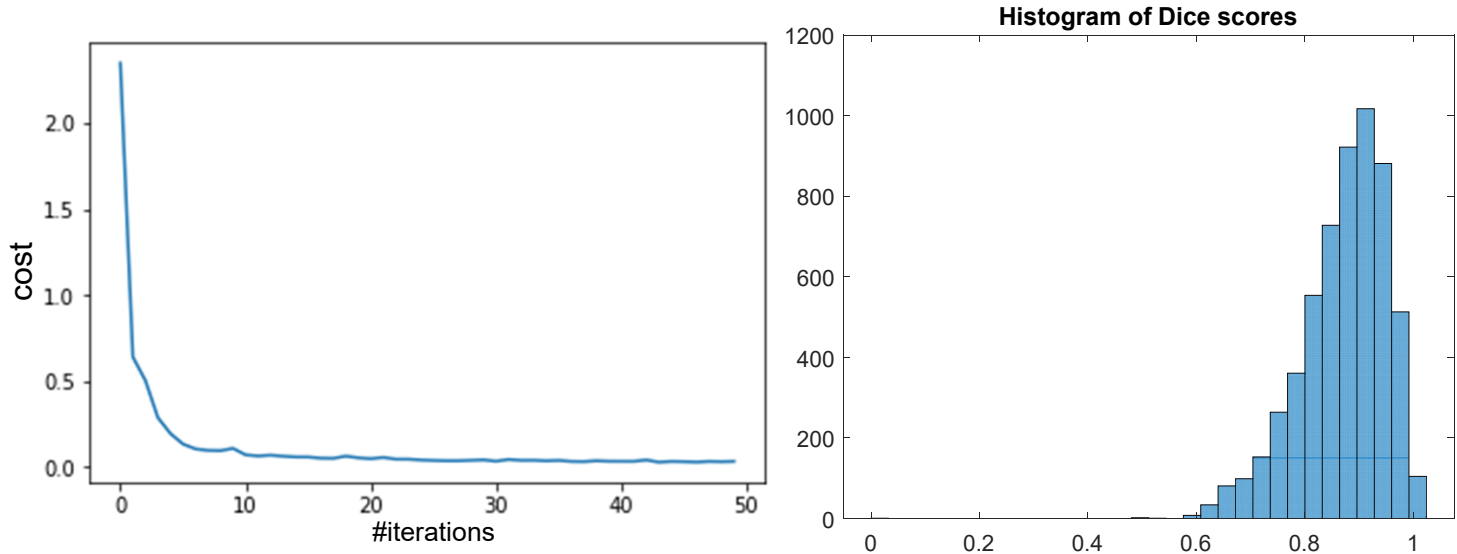

**Figure 4** AIF detection performance on the test set. (a) cost vs. the number of iterations for a typical training session, (b) histogram of three-class detection for all test cases, showing majority of cases have high dice scores.

(a) Cost vs. number of iterations

(b) Histogram of Dice ratios

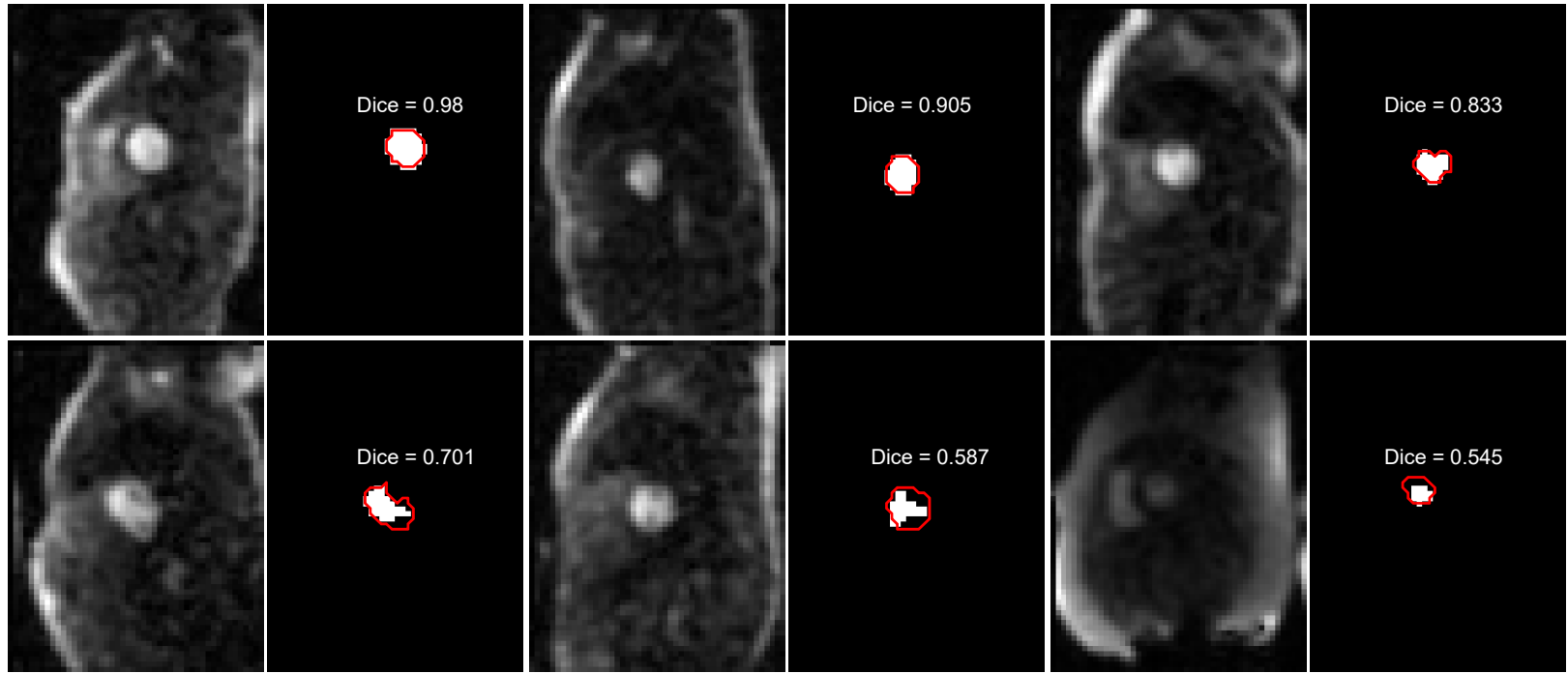

**Figure 5** Examples of AIF detection with different level of performance. Cases with lower score still located the LV blood pool successfully.

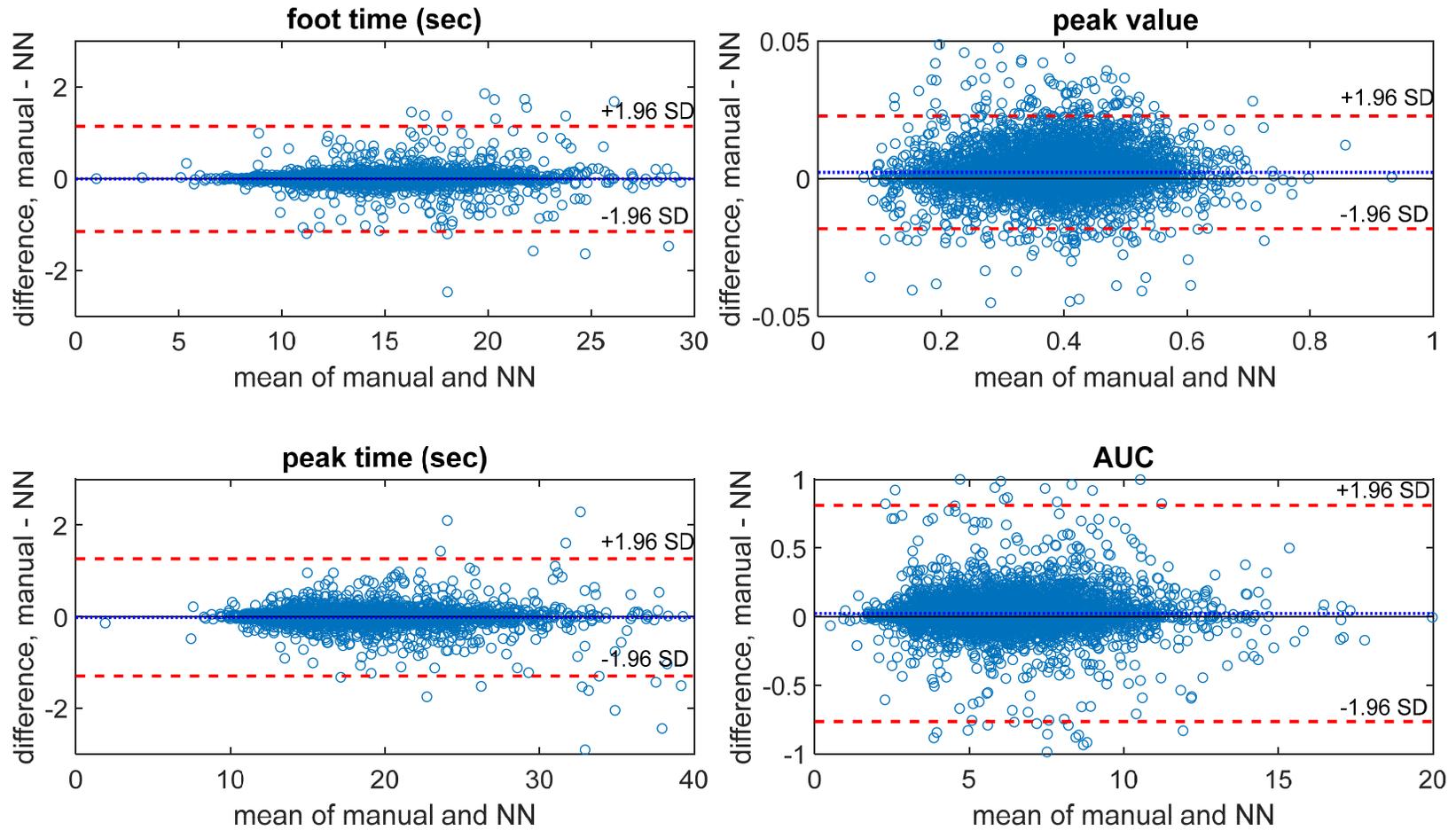

**Figure 6** Bland Altman plots for parameters derived from AIF signal curve comparing manually labelled vs. neural net based segmentation approaches. Both bias (dotted line in blue) and 95% limits of agreement (dotted lines in red) are plotted.

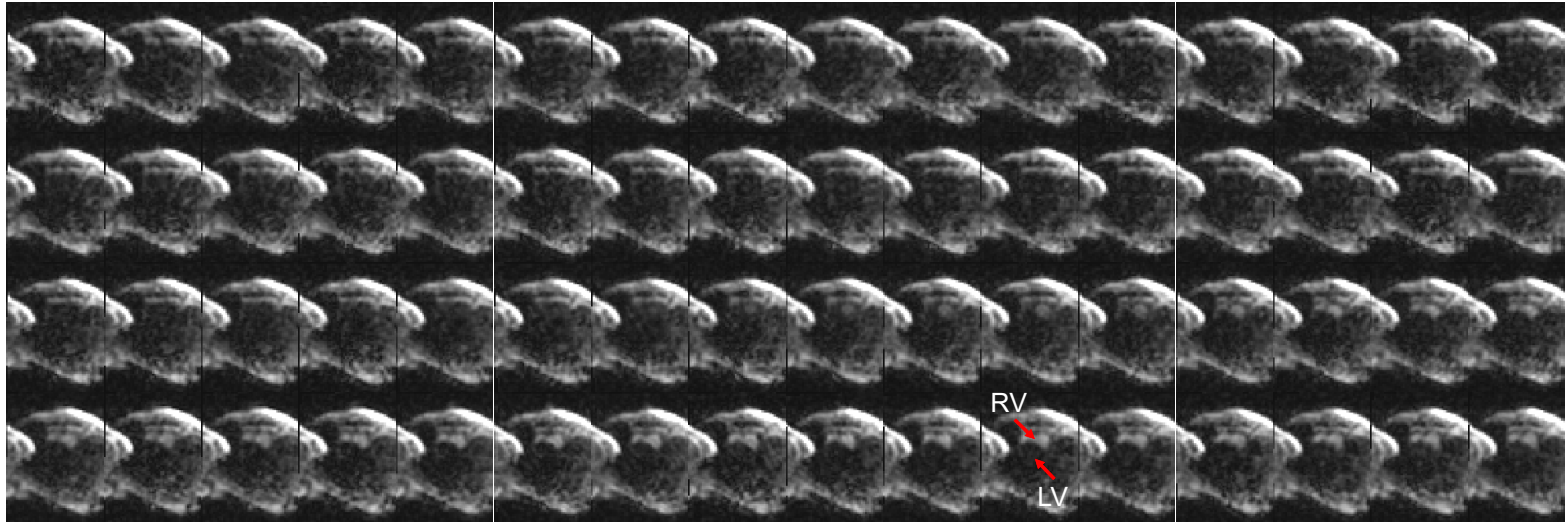

(a) AIF image series for the failed LV detection

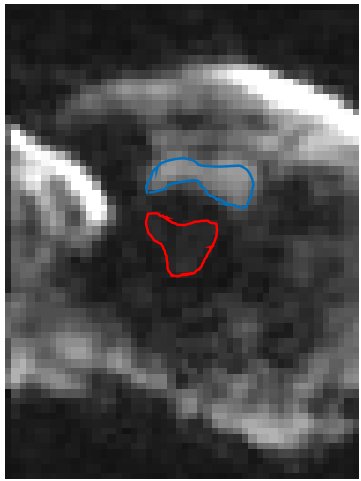

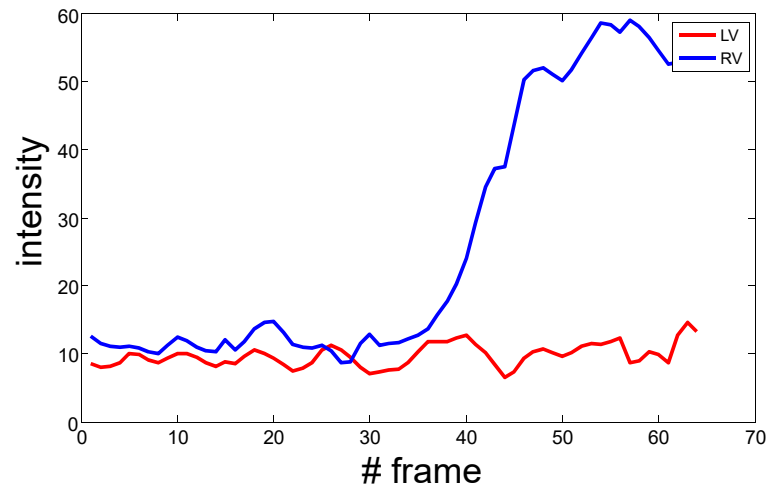

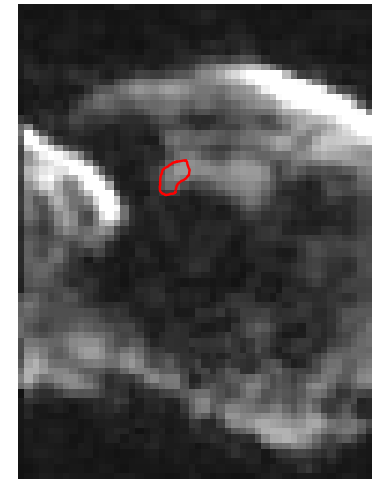

(b) Ground-truth masks for LV (red) and RV (blue)

(c) Time-intensity profiles for LV and RV

(d) Model incorrectly picked RV

**Supporting Information Figure S1** The AIF signal for the single failed case shows that the contrast did not reach the LV. (a) AIF time series is presented for all 64 frames. In this case, the contrast arrived very late and the LV did not enhance. (b) Manual labels for LV and RV are overlaid on a frame. (c) Time-intensity profiles for manually labelled ROIs show very late contrast arrival. (d) Neural network model mistakenly detected the RV.

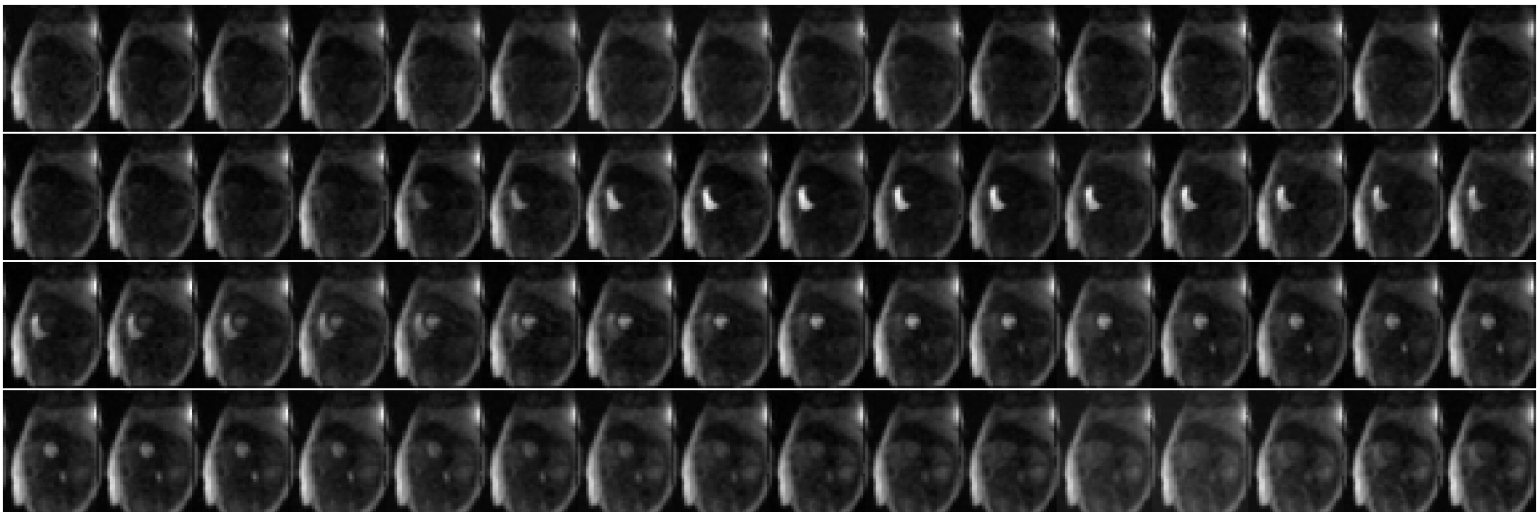

(a) An AIF image array to compute saliency map which is the derivative of loss function to input image array.

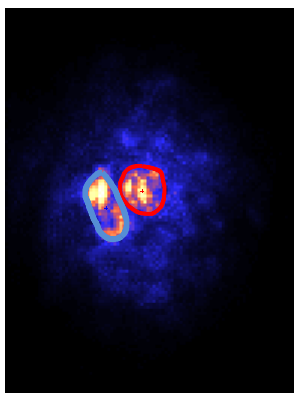 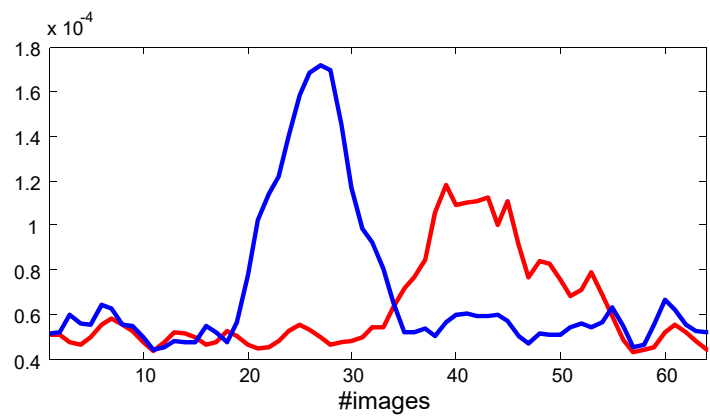

(b) The saliency map plotted as the maximum across temporal frames. Higher value in LV and RV indicates these regions have more impact on the final segmentation accuracy and the trained CNN is "paying attention" to these regions.

(c) The temporal profile of saliency response was plotted for RV(red) and LV(blue). The contrast bolus uptake led to elevated response in saliency map, indicating trained CNN responded more to the bolus passage period in time.

**Supporting Information Figure S2** The saliency map was computed for an AIF image array to investigate the characteristics of trained CNN. The saliency map was the derivative of loss function to input image array. Higher magnitude in saliency map indicated corresponding spatial and/or temporal regions in image had more effects on CNN loss.